\documentclass{aa}
\usepackage{color}
\usepackage{graphicx}
\usepackage{lscape}
\usepackage{natbib}
\usepackage[scaled]{helvet}
\usepackage[varg]{txfonts}
\usepackage{url}
\usepackage{xspace}
\bibpunct{(}{)}{;}{a}{}{,}
\newcounter{Rco}
\newcommand{\ionw}[3]{\mbox{\ion{#1}{#2}~$\lambda\,#3\,\mathrm{\AA}$}\xspace}
\newcommand{\ionww}[3]{\mbox{\ion{#1}{#2}~$\lambda\lambda\,#3\,\mathrm{\AA}$}\xspace}

\newcommand{\loggw}[1]{\mbox{$\log g\hspace{-0.5mm} =\hspace{-0.5mm}  #1$}}

\newcommand{\Teff}{\mbox{$T_\mathrm{eff}$}\xspace}
\newcommand{\Teffw}[1]{\mbox{$\Teff\hspace{-0.5mm} =\hspace{-0.5mm} #1 \,\mathrm{K}$}}

\newcommand{\mmspr}{\hbox{}\hspace{+0.8cm}}
\newcommand{\smspr}{\hbox{}\hspace{+2.5mm}}

\newcommand{\gb}{\object{G191$-$B2B}\xspace}
\newcommand{\re}{\object{RE\,0503$-$289}\xspace}
\begin{document}
\title{Stellar laboratories}
\subtitle{IV. New \ion{Ga}{iv}, \ion{Ga}{v}, and \ion{Ga}{vi} oscillator strengths and the gallium abundance \\
              in the hot white dwarfs \gb and \re
           \thanks
           {Based on observations with the NASA/ESA Hubble Space Telescope, obtained at the Space Telescope Science 
            Institute, which is operated by the Association of Universities for Research in Astronomy, Inc., under 
            NASA contract NAS5-26666.
           }$^,$
           \thanks
           {Based on observations made with the NASA-CNES-CSA Far Ultraviolet Spectroscopic Explorer.
           }$^,$
           \thanks
           {Tables 7 to 9 are only available at the CDS via anonymous ftp to
            cdsarc.u-strasbg.fr (130.79.128.5) or via
            http://cdsarc.u-strasbg.fr/viz-bin/qcat?J/A+A/vol/page
           }
         }
\titlerunning{Stellar laboratories: new \ion{Ga}{iv}, \ion{Ga}{v}, and \ion{Ga}{vi} oscillator strengths}

\author{T\@. Rauch\inst{1}
        \and
        K\@. Werner\inst{1}
        \and
        P\@. Quinet\inst{2,3}
        \and
        J\@. W\@. Kruk\inst{4}
        }

\institute{Institute for Astronomy and Astrophysics,
           Kepler Center for Astro and Particle Physics,
           Eberhard Karls University,
           Sand 1,
           72076 T\"ubingen,
           Germany \\
           \email{rauch@astro.uni-tuebingen.de}
           \and
           Astrophysique et Spectroscopie, Universit\'e de Mons -- UMONS, 7000 Mons, Belgium
           \and
           IPNAS, Universit\'e de Li\`ege, Sart Tilman, 4000 Li\`ege, Belgium
           \and
           NASA Goddard Space Flight Center, Greenbelt, MD\,20771, USA}

\date{Received 13 November 2014; accepted January 27 2015}

\abstract {For the spectral analysis of high-resolution and high-signal-to-noise (S/N) spectra of hot stars,
           advanced non-local thermodynamic equilibrium (NLTE) 
           model atmospheres are mandatory. These atmospheres are strongly
           dependent on the reliability of the atomic data that are used to calculate them.
          }
          {Reliable \ion{Ga}{iv-vi} oscillator strengths 
           are used to identify Ga lines in the spectra of 
           the DA-type white dwarf \gb and 
           the DO-type white dwarf \re and
           to determine their photospheric Ga abundances.
          }
          {We newly calculated \ion{Ga}{iv-vi} oscillator strengths
           to consider their radiative and collisional bound-bound transitions
           in detail in our NLTE stellar-atmosphere models
           for analyzing of Ga lines exhibited in
           high-resolution and high-S/N UV observations of \gb  and \re.
          }
          {We unambiguously detected 20 isolated and 6 blended (with lines of other
           species) \ion{Ga}{v} lines in the 
           Far Ultraviolet Spectroscopic Explorer (FUSE) spectrum of \re. 
           The identification of \ion{Ga}{iv} and \ion{Ga}{vi} lines is uncertain
           because they are weak and partly blended by other lines.
           The determined Ga abundance is $3.5 \pm 0.5 \times 10^{-5}$
           (mass fraction, about 625 times the solar value).
           The \ion{Ga}{iv}\,/\,\ion{Ga}{v} ionization equilibrium, which is a very sensitive indicator
           for the effective temperature, is well reproduced in \re. 
           We identified the strongest \ion{Ga}{iv} lines (at 1258.801, 1338.129\,\AA)
           in the HST/STIS (Hubble Space Telescope / Space Telescope Imaging Spectrograph) spectrum of \gb
           and measured a Ga abundance of $2.0 \pm 0.5 \times 10^{-6}$ (about 22 times solar).
           }
          {Reliable measurements and calculations of atomic data are a prerequisite for
           stellar-atmosphere modeling. 
           The observed \ion{Ga}{iv-v} line profiles in two white dwarf (\gb and \re) 
           ultraviolet spectra were well reproduced with our newly calculated oscillator strengths. 
           For the first time, this allowed us to determine the photospheric Ga abundance in white dwarfs.
          }


\maketitle

\section{Introduction}
\label{sect:intro}

The spectral lines of ten trans-iron elements were detected by \citet{werneretal2012} in the hydrogen-deficient 
DO-type white dwarf (WD) \re. The authors used Far Ultraviolet Spectroscopic Explorer (FUSE) spectra with a 
high resolution and high signal-to-noise ratio (S/N). 
\re has an effective temperature of \Teffw{70\,000} and a surface gravity of $\log\,(g\,/\,\mathrm{cm/s^2}) = 7.5$.
\citep{dreizlerwerner1996}. Ga and Mo were identified for the first time in a WD.
An abundance analysis was performed by \citet{werneretal2012} for
Kr and Xe ($-4.3 \pm 0.5$ and $-4.2 \pm 0.6$ in logarithmic mass fractions, respectively) alone
because they lacked atomic data for the other newly identified, highly ionized species.

New calculations of reliable transition probabilities (not only for the 
identified lines themselves, but for the complete model atom that is considered in 
the model atmosphere and spectral energy distribution (SED) calculations) 
are mandatory for precise abundance analyses. These enabled us to measure the
Zn \citep{rauchetal2014zn},
Ge \citep{rauchetal2012ge}, and 
Ba \citep{rauchetal2014ba} abundances in \re and the hydrogen-rich DA-type WD 
\gb \citep[\Teffw{60\,000}, \loggw{7.6},][]{rauchetal2013}. The Sn abundance in \gb was
determined based on existing atomic data \citep{rauchetal2013}.
Our models strongly reduced the number of unidentified lines in the spectra of \re and \gb.

Since \citet{werneretal2012} identified more than a dozen \ion{Ga}{v} lines in the
FUSE spectrum of \re, we computed \ion{Ga}{iv-vi} transition probabilities 
(Sect.\,\ref{sect:gatrans}) and calculated
non-local thermodynamic equilibrium (NLTE) model-atmosphere spectra (Sect.\,\ref{sect:models})
for an accurate Ga-abundance determination (Sect.\,\ref{sect:abund}). 
We summarize our results and conclude in Sect.\,\ref{sect:results}.

\section{Atomic structure and radiative data calculation}
\label{sect:gatrans}

Many spectral lines of \ion{Ga}{iv}, \ion{Ga}{v}, and \ion{Ga}{vi} were observed in
laboratories in the past. This allowed identifying very many energy levels of
\ion{Ga}{iv}, \ion{Ga}{v}, and \ion{Ga}{vi} that were listed by \citet{shiraietal2007},
who made a critical compilation of all the experimental data previously published by 
\citet{macketal1928}, \citet{moore1971}, \citet{ryabtsev1975}, \citet{ramonasryabtsev1990}, and \citet{ryabtsevchurilov1991} for \ion{Ga}{iv}, 
\citet{sawyerhumphreys1928}, \citet{kononov1967}, \citet{joshietal1972}, \citet{aksenovryabtsev1974}, \citet{dick1974}, 
\citet{vandeurzen1977}, and \citet{ryabtsevramonas1985} for \ion{Ga}{v}, and 
\citet{podobedovaetal1983,podobedovaetal1985} for \ion{Ga}{vi}.
   
In the present work, new sets of oscillator strengths and transition probabilities were obtained  for \ion{Ga}{iv}, \ion{Ga}{v}, and \ion{Ga}{vi}. 
These were computed using the pseudo-relativistic Hartree-Fock (HFR) approach of \citet{cowan1981} combined with a semi-empirical 
least-squares fit of radial energy parameters. In each ion, many electron correlations were considered by means of 
extended multiconfiguration expansions that are included in the physical models. These expansions were chosen so as to include low-lying 
configurations for which energy levels are experimentally known together with some higher configurations with large configuration 
interaction Slater integrals $R^\mathrm{k}$ that connect these latter configurations to the former.

More precisely, in \ion{Ga}{iv}, configuration interaction was explicitly considered among the configurations 
3d$^{10}$, 
3d$^{9}$4s, 
3d$^{9}$5s, 
3d$^{9}$6s, 
3d$^{9}$7s, 
3d$^{9}$4d, 
3d$^{9}$5d, 
3d$^{9}$6d, 
3d$^{9}$7d, 
3d$^{8}$4s$^{2}$, 
3d$^{8}$4p$^{2}$, 
3d$^{8}$4d$^{2}$, 
3d$^{8}$4f$^{2}$, 
3d$^{8}$4s5s, 
3d$^{8}$4s6s, 
3d$^{8}$4s7s, 
3d$^{8}$4s4d, 
3d$^{8}$4s5d, 
3d$^{8}$4s6d, 
3d$^{8}$4s7d, and 
3d$^{8}$4p4f for the even parity, and 
3d$^{9}$4p, 
3d$^{9}$5p, 
3d$^{9}$6p, 
3d$^{9}$7p, 
3d$^{9}$4f, 
3d$^{9}$5f, 
3d$^{9}$6f, 
3d$^{9}$7f, 
3d$^{8}$4s4p, 
3d$^{8}$4s5p, 
3d$^{8}$4s6p, 
3d$^{8}$4s7p, 
3d$^{8}$4s4f, 
3d$^{8}$4s5f, 
3d$^{8}$4s6f, 
3d$^{8}$4s7f, and 
3d$^{8}$4p4d for the odd parity. 
Using experimental energy levels reported by \citet{shiraietal2007}, the radial integrals (average energy, Slater, 
spin-orbit parameters and effective interaction parameters) of 
3d$^{10}$, 
3d$^{9}$4s, 
3d$^{9}$5s, 
3d$^{9}$6s, 
3d$^{9}$7s, 
3d$^{9}$4d, 
3d$^{9}$5d, and 
3d$^{9}$6d for even configurations and 
3d$^{9}$4p, 
3d$^{9}$5p, 
3d$^{9}$6p, 
3d$^{9}$4f, 
3d$^{8}$4s4p for odd configurations were optimized by a well-established fitting procedure. The mean deviations between computed and 
experimental energy levels were 61\,cm$^{-1}$ 
(71 levels)
 and 52\,cm$^{-1}$ 
(116 levels)
 for even and odd parities, respectively.

For \ion{Ga}{v}, the HFR method was used with, as interacting configurations, 
3d$^{9}$, 
3d$^{8}$4s, 
3d$^{8}$5s, 
3d$^{8}$4d, 
3d$^{8}$5d, 
3d$^{7}$4s$^{2}$, 
3d$^{7}$4p$^{2}$, 
3d$^{7}$4d$^{2}$, 
3d$^{7}$4f$^{2}$, 
3d$^{7}$4s5s, 
3d$^{7}$4s4d, and 
3d$^{7}$4s5d for the even parity, and 
3d$^{8}$4p, 
3d$^{8}$5p, 
3d$^{8}$4f, 
3d$^{8}$5f, 
3d$^{7}$4s4p, 
3d$^{7}$4s5p, 
3d$^{7}$4s4f, 
3d$^{7}$4s5f, and 
3d$^{7}$4p4d for the odd parity. The radial integrals corresponding to 
3d$^{9}$, 
3d$^{8}$4s, 
3d$^{8}$4p, and 
3d$^{8}$4f were adjusted to minimize the differences between the calculated Hamiltonian eigenvalues and the experimental energy levels 
taken from \citet{shiraietal2007}. In this process, we found mean deviations equal to 34\,cm$^{-1}$ 
(17 levels)
in the even parity and 108\,cm$^{-1}$ 
(74 levels)
in the odd parity.

Finally, in the case of \ion{Ga}{vi}, the configurations included in the HFR model were 
3d$^{8}$, 
3d$^{7}$4s, 
3d$^{7}$4d, 
3d$^{6}$4s$^{2}$, 
3d$^{6}$4p$^{2}$, 
3d$^{6}$4d$^{2}$, and 
3d$^{6}$4s4d for the even parity, and 
3d$^{7}$4p, 
3d$^{7}$4f, 
3d$^{6}$4s4p, 
3d$^{6}$4s4f, and 
3d$^{6}$4p4d for the odd parity. The experimental energy levels reported by \citet{shiraietal2007} were used 
here as well to optimize the 
radial integrals characterizing the 
3d$^{8}$, 
3d$^{7}$4s, and 
3d$^{7}$4p configurations. This semi-empirical process led to average deviations with experimental data equal to 122\,cm$^{-1}$ 
(47 levels)
and 161\,cm$^{-1}$ 
(110 levels)
for even and odd parities, respectively.

The numerical values of the parameters adopted in the present calculations are reported in 
Tables \ref{tab:gaiv:para}, \ref{tab:gav:para}, and \ref{tab:gavi:para}, while the computed energies are compared with available experimental values in 
Tables \ref{tab:gaiv:ener}, \ref{tab:gav:ener}, and \ref{tab:gavi:ener}, for \ion{Ga}{iv}, \ion{Ga}{v}, and \ion{Ga}{vi}, respectively.
Tables \ref{tab:gaiv:loggf}, \ref{tab:gav:loggf}, and \ref{tab:gavi:loggf} give the HFR oscillator strengths 
($\log gf$) and transition probabilities ($gA$, in s$^{-1}$) for \ion{Ga}{iv-vi}, respectively, and  
the numerical values (in cm$^{-1}$) of lower and upper energy levels and the corresponding wavelengths (in \AA). 
In the last column of each table, we also give absolute value of the cancellation factor CF as defined by \citet{cowan1981}. 
We note that very low values of this  factor (typically $< 0.05$) indicate strong cancellation effects in the
calculation of line strengths. In these cases, the corresponding $gf$ and $gA$ values could be very inaccurate 
and therefore need to be considered with some care. However, very few of the transitions appearing in 
Tables \ref{tab:gaiv:loggf},  \ref{tab:gav:loggf}, and \ref{tab:gavi:loggf} are affected.

\onltab{
\onecolumn
\begin{longtable}{llrrrl}
\caption{\label{tab:gaiv:para}Radial parameters (in cm$^{-1}$) adopted for the calculations in \ion{Ga}{iv}.} \\
\hline\hline
\noalign{\smallskip}
Configuration & Parameter & HFR & Fitted & Ratio & Note$^a$ \\ 
\noalign{\smallskip}
\hline
\endfirsthead
\caption{continued.}\\
\hline\hline
\noalign{\smallskip}
Configuration & Parameter & HFR & Fitted & Ratio & Note$^a$ \\
\noalign{\smallskip}
\hline
\noalign{\smallskip}
\endhead
\hline
\noalign{\smallskip}
\endfoot
\noalign{\smallskip}
\multicolumn{6}{c}{Even parity} \\
3d$^{10}$        & E$_{av}$          &    5814 &   5776 &       &    \\
3d$^9$4s         & E$_{av}$          &  148422 & 156111 &       &    \\
                 & $\zeta$$_{3d}$    &    1407 &   1429 & 1.016 &    \\
                 & G$^2$(3d,4s)      &   12329 &  11238 & 0.912 &    \\
3d$^9$5s         & E$_{av}$          &  339474 & 345159 &       &    \\
                 & $\zeta$$_{3d}$    &    1417 &   1428 & 1.008 &    \\
                 & G$^2$(3d,5s)      &    2908 &   2788 & 0.959 &    \\
\hline
\end{longtable}
$^{(a)}$ F: parameter fixed to its \emph{ab initio} value; R1, R2, R3, R4: ratios of these parameters were fixed in the fitting process.\\
\noindent
\twocolumn
}

\onltab{
\onecolumn
\begin{longtable}{llrrrl}
\caption{\label{tab:gav:para}Same as Table\,\ref{tab:gaiv:para} for \ion{Ga}{v}.} \\
\hline\hline
\noalign{\smallskip}
Configuration & Parameter & HFR & Fitted & Ratio & Note$^a$ \\ 
\noalign{\smallskip}
\hline
\endfirsthead
\caption{continued.}\\
\hline\hline
\noalign{\smallskip}
Configuration & Parameter & HFR & Fitted & Ratio & Note$^a$ \\
\noalign{\smallskip}
\hline
\noalign{\smallskip}
\endhead
\hline
\noalign{\smallskip}
\endfoot
\noalign{\smallskip}
\multicolumn{6}{c}{Even parity} \\
3d$^{9}$         & E$_{av}$          &    5908 &   5922 &       &    \\
                 & $\zeta$$_{3d}$    &    1420 &   1434 & 1.010 &    \\
3d$^8$4s         & E$_{av}$          &  226836 & 232051 &       &    \\
                 & F$^2$(3d,3d)      &  135437 & 119585 & 0.883 &    \\ 
                 & F$^4$(3d,3d)      &   85127 &  72403 & 0.851 &    \\
                 & $\alpha$          &       0 &    114 &       &    \\
                 & $\zeta$$_{3d}$    &    1518 &   1540 & 1.015 &    \\
\hline
\end{longtable}
\noindent
$^{(a)}$ F: parameter fixed to its \emph{ab initio} value; R1, R2: ratios of these parameters were fixed in the fitting process.\\
\twocolumn
}

\onltab{
\onecolumn
\begin{longtable}{llrrr}
\caption{\label{tab:gavi:para}Same as Table\,\ref{tab:gaiv:para} for \ion{Ga}{vi}.} \\
\hline\hline
\noalign{\smallskip}
Configuration & Parameter & HFR & Fitted & Ratio \\ 
\noalign{\smallskip}
\hline
\endfirsthead
\caption{continued.}\\
\hline\hline
\noalign{\smallskip}
Configuration & Parameter & HFR & Fitted & Ratio \\
\noalign{\smallskip}
\hline
\noalign{\smallskip}
\endhead
\hline
\noalign{\smallskip}
\endfoot
\noalign{\smallskip}
\multicolumn{5}{c}{Even parity} \\
3d$^8$           & E$_{av}$          &   20524 &  20236 &           \\
                 & F$^2$(3d,3d)      &  136668 & 121986 & 0.893     \\ 
                 & F$^4$(3d,3d)      &   85965 &  74905 & 0.871     \\
                 & $\alpha$          &       0 &    113 &           \\
                 & $\beta$           &       0 &   -418 &           \\
                 & $\zeta$$_{3d}$    &    1534 &   1548 & 1.009     \\
3d$^7$4s         & E$_{av}$          &  327549 & 330704 &           \\
\hline
\end{longtable}
\twocolumn
}

\onltab{
\onecolumn
\begin{longtable}{rrrrl}
\caption{\label{tab:gaiv:ener} Comparison between experimental and calculated energy levels in \ion{Ga}{iv}. Energies are given in cm$^{-1}$.} \\
\hline\hline
\noalign{\smallskip}
E$_{exp}$$^a$  & E$_{calc}$$^b$ & $\Delta$E & J & Leading components (in \%) in $LS$ coupling$^c$ \\ 
\noalign{\smallskip}
\hline
\endfirsthead
\caption{continued.}\\
\hline\hline
\noalign{\smallskip}
E$_{exp}$$^a$  & E$_{calc}$$^b$ & $\Delta$E & J & Leading components (in \%) in $LS$ coupling$^c$ \\
\noalign{\smallskip}
\hline
\noalign{\smallskip}
\endhead
\hline
\noalign{\smallskip}
\endfoot
\noalign{\smallskip}
\multicolumn{5}{c}{Even parity} \\

        0.0 &        0 &     0 & 0 &   99 3d$^{10}$ $^1$S \\   
   149512.1 &   149518 &  $-$6 & 3 &   99 3d$^9$4s $^3$D \\     
   150967.0 &   150959 &     8 & 2 &   85 3d$^9$4s $^3$D + 14 3d$^9$4s $^1$D \\      
   153085.9 &   153087 &  $-$1 & 1 &   99 3d$^9$4s $^3$D \\         
   156024.7 &   156026 &  $-$1 & 2 &   85 3d$^9$4s $^1$D + 14 3d$^9$4s $^3$D \\      
   328763.7 &   328928 &$-$164 & 1 &   91 3d$^9$4d $^3$S + 6 3d$^9$4d $^3$P \\      
   332554.3 &   332406 &   149 & 5 &   99 3d$^9$4d $^3$G \\
\hline
\end{longtable}
\noindent
$^{(a)}$ From \citet{shiraietal2007}.\\
$^{(b)}$ This work.\\
$^{(c)}$ Only the first three components that are larger than 5\% are given.
\twocolumn
}

\onltab{
\onecolumn
\begin{longtable}{rrrrl}
\caption{\label{tab:gav:ener} Same as Table\,\ref{tab:gavi:ener} for \ion{Ga}{v}.} \\
\hline\hline
\noalign{\smallskip}
E$_{exp}$$^a$  & E$_{calc}$$^b$ & $\Delta$E & J & Leading components (in \%) in $LS$ coupling$^c$ \\ 
\noalign{\smallskip}
\hline
\endfirsthead
\caption{continued.}\\
\hline\hline
\noalign{\smallskip}
E$_{exp}$$^a$  & E$_{calc}$$^b$ & $\Delta$E & J & Leading components (in \%) in $LS$ coupling$^c$ \\
\noalign{\smallskip}
\hline
\noalign{\smallskip}
\endhead
\hline
\noalign{\smallskip}
\endfoot
\noalign{\smallskip}
\multicolumn{5}{c}{Even parity} \\
        0 &        0 &     0 & 2.5 &   99  3d$^9$   $^2$D \\   
     3583 &     3583 &     0 & 1.5 &   99  3d$^9$   $^2$D \\  
   210052 &   210053 &  $-$1 & 4.5 &   99  3d$^8$($^3$F)4s $^4$F \\     
   212121 &   212098 &    23 & 3.5 &   95  3d$^8$($^3$F)4s $^4$F \\     
   214000 &   213960 &    40 & 2.5 &   97  3d$^8$($^3$F)4s $^4$F \\
   215237 &   215192 &    45 & 1.5 &   98  3d$^8$($^3$F)4s $^4$F \\  
   218301 &   218359 & $-$58 & 3.5 &   94  3d$^8$($^3$F)4s $^2$F  +   5  3d$^8$($^3$F)4s $^4$F \\
\hline
\end{longtable}
\noindent
$^{(a)}$ From \citet{shiraietal2007}.\\
$^{(b)}$ This work.\\
$^{(c)}$ Only the first three components that are larger than 5\% are given.
\twocolumn
}

\onltab{
\onecolumn
\begin{longtable}{rrrrl}
\caption{\label{tab:gavi:ener}Same as Table\,\ref{tab:gavi:ener} for \ion{Ga}{vi}.} \\
\hline\hline
\noalign{\smallskip}
E$_{exp}$$^a$  & E$_{calc}$$^b$ & $\Delta$E & J & Leading components (in \%) in $LS$ coupling$^c$ \\ 
\noalign{\smallskip}
\hline
\endfirsthead
\caption{continued.}\\
\hline\hline
\noalign{\smallskip}
E$_{exp}$$^a$  & E$_{calc}$$^b$ & $\Delta$E & J & Leading components (in \%) in $LS$ coupling$^c$ \\
\noalign{\smallskip}
\hline
\noalign{\smallskip}
\endhead
\hline
\noalign{\smallskip}
\endfoot
\noalign{\smallskip}
\multicolumn{5}{c}{Even parity} \\
        0.0 &       23 &  $-$23 & 4 &   99  3d$^8$ $^3$F \\      
     3193.3 &     3187 &      7 & 3 &  100  3d$^8$ $^3$F \\            
     5164.0 &     5147 &     17 & 2 &   97  3d$^8$ $^3$F \\            
    20531.6 &    20507 &     24 & 2 &   75  3d$^8$ $^1$D  +  23  3d$^8$ $^3$P \\           
    25683.2 &    25765 &  $-$82 & 2 &   76  3d$^8$ $^3$P  +  23  3d$^8$ $^1$D \\           
    26123.7 &    26090 &     33 & 1 &   99  3d$^8$ $^3$P \\           
    26607.4 &    26583 &     24 & 0 &   99  3d$^8$ $^3$P \\
\hline
\end{longtable}
\noindent
$^{(a)}$ From \citet{shiraietal2007}.\\
$^{(b)}$ This work.\\
$^{(c)}$ Only the first three components that are larger than 5\% are given.
\twocolumn
}

\onltab{
\onecolumn
\begin{longtable}{rrccrrccrrr}
\caption{\label{tab:gaiv:loggf}Calculated HFR oscillator strengths ($\log gf$) and transition probabilities (gA) 
                              in \ion{Ga}{iv}. CF is the cancellation factor as defined by \citet{cowan1981}. 
                              In Cols\@. 3 and 6, e is written for even and o for odd.} \\
\hline\hline
\noalign{\smallskip}
                 & \multicolumn{3}{c}{Lower level} && \multicolumn{3}{c}{Upper level} &           &      &    \\
\cline{2-4}\cline{6-8}
\noalign{\smallskip}
Wavelength{$^a$} / \AA & Energy{$^b$} / cm$^{-1}$ & Parity & $J$  && Energy{$^b$} / cm$^{-1}$ & Parity & $J$  & $\log gf$ & $gA$ / s$^{-1}$  & CF \\
\noalign{\smallskip}
\hline
\endfirsthead
\caption{continued.}\\
\hline\hline
\noalign{\smallskip}
                 & \multicolumn{3}{c}{Lower level} && \multicolumn{3}{c}{Upper level} &           &      &    \\
\cline{2-4}\cline{6-8}
\noalign{\smallskip}
Wavelength{$^a$} / \AA & Energy{$^b$} / cm$^{-1}$ & Parity & $J$  && Energy{$^b$} / cm$^{-1}$ & Parity & $J$  & $\log gf$ & $gA$ / s$^{-1}$  & CF \\
\noalign{\smallskip}
\hline
\noalign{\smallskip}
\endhead
\hline
\noalign{\smallskip}
\endfoot
\noalign{\smallskip}
     213.777 &         0 & e & 0.0 & &    467778 & o & 1.0 &$-$2.69 &  $3.01 \times 10^{08} $ & 0.034 \\
     215.415 &         0 & e & 0.0 & &    464221 & o & 1.0 &$-$2.18 &  $9.47 \times 10^{08} $ & 0.082 \\
     225.558 &         0 & e & 0.0 & &    443344 & o & 1.0 &$-$1.84 &  $1.91 \times 10^{09} $ & 0.141 \\
     227.155 &         0 & e & 0.0 & &    440229 & o & 1.0 &$-$1.75 &  $2.31 \times 10^{09} $ & 0.325 \\
     227.588 &         0 & e & 0.0 & &    439390 & o & 1.0 &$-$2.20 &  $8.11 \times 10^{08} $ & 0.229 \\
     232.778 &         0 & e & 0.0 & &    429594 & o & 1.0 &$-$2.82 &  $1.88 \times 10^{08} $ & 0.076 \\
     233.099 &         0 & e & 0.0 & &    429003 & o & 1.0 &$-$1.83 &  $1.83 \times 10^{09} $ & 0.085 \\
     234.823 &         0 & e & 0.0 & &    425853 & o & 1.0 &$-$1.86 &  $1.67 \times 10^{09} $ & 0.094 \\
\end{longtable}
\twocolumn
}

\onltab{
\onecolumn
\begin{longtable}{rrccrrccrrr}
\caption{\label{tab:gav:loggf}
Same as Table\,\ref{tab:gaiv:loggf}, for \ion{Ga}{v}.} \\
\hline\hline
\noalign{\smallskip}
                 & \multicolumn{3}{c}{Lower level} && \multicolumn{3}{c}{Upper level} &           &      &    \\
\cline{2-4}\cline{6-8}
\noalign{\smallskip}
Wavelength{$^a$} / \AA & Energy{$^b$} / cm$^{-1}$ & Parity & $J$  && Energy{$^b$} / cm$^{-1}$ & Parity & $J$  & $\log gf$ & $gA$ / s$^{-1}$  & CF \\
\noalign{\smallskip}
\hline
\endfirsthead
\caption{continued.}\\
\hline\hline
\noalign{\smallskip}
                 & \multicolumn{3}{c}{Lower level} && \multicolumn{3}{c}{Upper level} &           &      &    \\
\cline{2-4}\cline{6-8}
\noalign{\smallskip}
Wavelength{$^a$} / \AA & Energy{$^b$} / cm$^{-1}$ & Parity & $J$  && Energy{$^b$} / cm$^{-1}$ & Parity & $J$  & $\log gf$ & $gA$ / s$^{-1}$  & CF \\
\noalign{\smallskip}
\hline
\noalign{\smallskip}
\endhead
\hline
\noalign{\smallskip}
\endfoot
\noalign{\smallskip}
     167.404 &         0 & e & 2.5 & &    597358 & o & 2.5 &$-$2.49 &  $7.68 \times 10^{08} $ & 0.151 \\
     167.456 &         0 & e & 2.5 & &    597172 & o & 3.5 &$-$0.99 &  $2.44 \times 10^{10} $ & 0.464 \\
     168.414 &      3583 & e & 1.5 & &    597358 & o & 2.5 &$-$0.99 &  $2.39 \times 10^{10} $ & 0.548 \\
     180.553 &         0 & e & 2.5 & &    553853 & o & 2.5 &$-$1.78 &  $3.43 \times 10^{09} $ & 0.484 \\
     180.563 &         0 & e & 2.5 & &    553823 & o & 3.5 &$-$0.82 &  $3.13 \times 10^{10} $ & 0.474 \\
     180.806 &         0 & e & 2.5 & &    553080 & o & 1.5 &$-$1.42 &  $7.72 \times 10^{09} $ & 0.592 \\
     180.832 &         0 & e & 2.5 & &    552998 & o & 2.5 &$-$0.44 &  $7.42 \times 10^{10} $ & 0.574 \\
     181.160 &         0 & e & 2.5 & &    551998 & o & 1.5 &$-$0.38 &  $8.42 \times 10^{10} $ & 0.645 \\
\end{longtable}
\twocolumn
}

\onltab{
\onecolumn
\begin{longtable}{rrccrrccrrr}
\caption{\label{tab:gavi:loggf}
Same as Table\,\ref{tab:gaiv:loggf}, for \ion{Ga}{vi}.} \\
\hline\hline
\noalign{\smallskip}
                 & \multicolumn{3}{c}{Lower level} && \multicolumn{3}{c}{Upper level} &           &      &    \\
\cline{2-4}\cline{6-8}
\noalign{\smallskip}
Wavelength{$^a$} / \AA & Energy{$^b$} / cm$^{-1}$ & Parity & $J$  && Energy{$^b$} / cm$^{-1}$ & Parity & $J$  & $\log gf$ & $gA$ / s$^{-1}$ & CF \\
\noalign{\smallskip}
\hline
\endfirsthead
\caption{continued.}\\
\hline\hline
\noalign{\smallskip}
                 & \multicolumn{3}{c}{Lower level} && \multicolumn{3}{c}{Upper level} &           &      &    \\
\cline{2-4}\cline{6-8}
\noalign{\smallskip}
Wavelength{$^a$} / \AA & Energy{$^b$} / cm$^{-1}$ & Parity & $J$  && Energy{$^b$} / cm$^{-1}$ & Parity & $J$  & $\log gf$ & $gA$ / s$^{-1}$ & CF \\
\noalign{\smallskip}
\hline
\noalign{\smallskip}
\endhead
\hline
\noalign{\smallskip}
\endfoot
\noalign{\smallskip}
     200.053 &         0 & e & 4.0 & &    499867 & o & 3.0 &$-$1.39 &  $6.72 \times 10^{09} $ & 0.081 \\
     201.339 &      3193 & e & 3.0 & &    499867 & o & 3.0 &$-$2.41 &  $6.44 \times 10^{08} $ & 0.071 \\
     201.724 &      3193 & e & 3.0 & &    498921 & o & 2.0 &$-$2.59 &  $4.19 \times 10^{08} $ & 0.053 \\
     202.143 &      3193 & e & 3.0 & &    497894 & o & 2.0 &$-$1.50 &  $5.18 \times 10^{09} $ & 0.104 \\
     202.529 &      5164 & e & 2.0 & &    498921 & o & 2.0 &$-$2.69 &  $3.31 \times 10^{08} $ & 0.031 \\
     202.649 &         0 & e & 4.0 & &    493465 & o & 3.0 &$-$2.77 &  $2.74 \times 10^{08} $ & 0.030 \\
     202.951 &      5164 & e & 2.0 & &    497894 & o & 2.0 &$-$2.53 &  $4.83 \times 10^{08} $ & 0.039 \\
     203.183 &      5164 & e & 2.0 & &    497330 & o & 1.0 &$-$1.64 &  $3.70 \times 10^{09} $ & 0.101 \\
\end{longtable}
\twocolumn
}

\section{Observations}
\label{sect:observation}

We analyzed the FUSE spectrum 
($910\,\mathrm{\AA} < \lambda <  1188\,\mathrm{\AA}$, resolving power $R = \lambda/\Delta\lambda \approx 20\,000$)
of \re and 
for \gb FUSE and Hubble Space Telescope / Space Telescope Imaging Spectrograph
spectra (HST/STIS, $1145\,\mathrm{\AA} < \lambda < 1750\,\mathrm{\AA}$).
The latter is co-added from 105 observations with the highest resolution 
(grating E140H, $R \approx$ $118\,000$, \url{http://www.stsci.edu/hst/observatory/crds/calspec.html}).
These spectra were previously described in detail by \citet{werneretal2012} and \citet{rauchetal2013}, respectively.

In addition, we used our recently obtained (and later co-added) HST/STIS spectra of \re 
(2014-08-14, ObsIds OC7N01010, OC7N01020, grating E140M, $R \approx 45\,800$,
total exposure time 5494\,s).

All spectra are available via the Barbara A\@. Mikulski Archive for Space Telescopes 
(MAST\footnote{\url{http://archive.stsci.edu/}}).

\section{Model atmospheres and atomic data}
\label{sect:models}

The T\"ubingen NLTE model-atmosphere package 
\citep[TMAP\footnote{\url{http://astro.uni-tuebingen.de/~TMAP}},][]{werneretal2003,rauchdeetjen2003}
was used to calculate advanced plane-parallel and chemically homogeneous stellar
model atmospheres in radiative and hydrostatic equilibrium.

In our recent abundance analyses of trans-iron elements \citep{rauchetal2014zn, rauchetal2014ba},
we encountered the problem that with a new species included in the models, our model-atmosphere code
would not compile if the array sizes were increased to account for the higher number of atomic levels 
treated in NLTE and the respective higher number of radiative and collisional transitions. 
Therefore, we decided at that time to reduce the number 
of levels of various species in our model atoms that were treated in NLTE.
This did not have a significant effect on the abundance analyses.
However, since we plan to continue the abundance analyses of trans-iron elements, an incalculable, systematic
error may arise. 

To represent the elements with an atomic number $<20$ with sufficient detail,
we took the prepared (``classical'') model atoms
that are provided by the T\"ubingen Model-Atom Database
\citep[TMAD\footnote{\url{http://astro.uni-tuebingen.de/~TMAD}},][]{rauchdeetjen2003}.
TMAD was constructed in the framework of the German Astrophysical Virtual Observatory 
(GAVO\footnote{\url{http://www.g-vo.org}}).

We decided to construct the model atoms for the trans-iron elements
similar to those of Ca to Ni \citep{rauchdeetjen2003}.
To reduce the number of atomic levels and spectral lines considered in our
model-atmosphere calculations, we employed our program
Iron Opacity and Interface \citep[IrOnIc,][]{rauchdeetjen2003}, which uses a statistical approach to calculate so-called
super levels and super lines. Table\,\ref{tab:ironic} demonstrates the
strongly extenuated level and line numbers for Ga.

\begin{table}\centering
\caption{\ion{Ga}{iv}, \ion{Ga}{v}, and \ion{Ga}{vi} atomic levels and line transitions from
         Tables\,\ref{tab:gaiv:loggf}, \ref{tab:gav:loggf}, and \ref{tab:gavi:loggf}, respectively.
         The super levels and super lines are calculated by IrOnIc \citep{rauchdeetjen2003}.}         
\label{tab:ironic}
\begin{tabular}{ccccc}
\hline
\hline
ion       & atomic levels & lines & super levels & super lines \\
\hline
\sc{iv}   &           191 &  3198 &            7 &          19 \\
\sc{v}    &            91 &   517 &            7 &          15 \\
\sc{vi}   &           157 &  1914 &            7 &          13 \\
\hline
          &           439 &  5629 &           21 &          47 \\
\hline
\end{tabular}
\end{table}

IrOnIc was designed to read data in the
Kurucz format\footnote{GFxxyy.GAM, GFxxyy.LIN, and GFxxyy.POS files with xx = element number,
yy = element charge, \url{http://kurucz.harvard.edu/atoms.html}}
\citep{kurucz1991} as well as in the 
Opacity Project format\footnote{\url{http://cdsweb.u-strasbg.fr/topbase/topbase.html}}.
We transferred our Zn, Ga, Ge, and Ba data into Kurucz-formatted files that were then ingested
and processed by IrOnIc. The ``statistical'' model atoms that are created by this means were then
used together with the ``classical'' model atoms for the lighter metals (see above) in our model 
atmosphere calculations.

\begin{figure}
   \resizebox{\hsize}{!}{\includegraphics{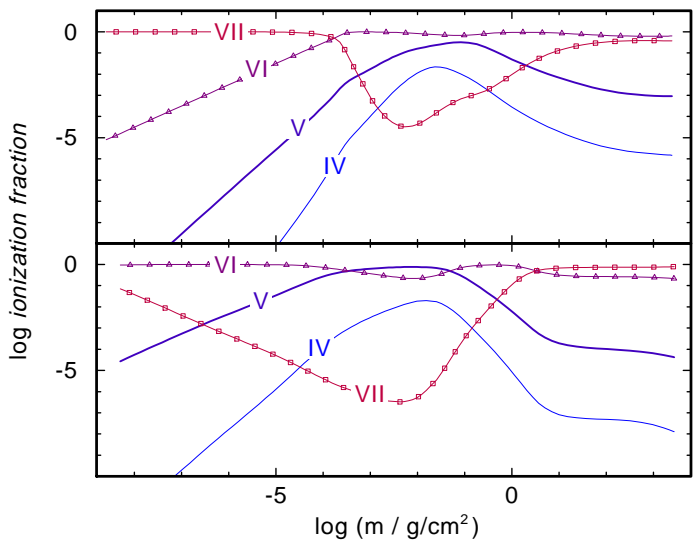}}
    \caption{Ga ionization fractions in our models for 
             \gb (top panel, H+Ga model) and
             \re (bottom panel, He+Ga).
            }
   \label{fig:ion}
\end{figure}

To verify the reliability of this approach, we calculated models that consider only H+Ga and He+Ga 
with a 100 times solar Ga abundance \citep[mass fraction $5.6\times10^{-6}$, solar value from ][]{asplundetal2009}
for
\gb (\Teffw{60\,000}, \loggw{7.6}) and 
\re (\Teffw{70\,000}, \loggw{7.5}).
Figure\,\ref{fig:ion} displays the Ga ionization fractions in these models.
\ion{Ga}{v-vi} are the dominant ionization stages in the line-forming region ($-4 \la \log\,m \la 0.5$.
$m$ is the column mass, measured from the outer boundary of our model atmospheres.

\begin{figure}
   \resizebox{\hsize}{!}{\includegraphics{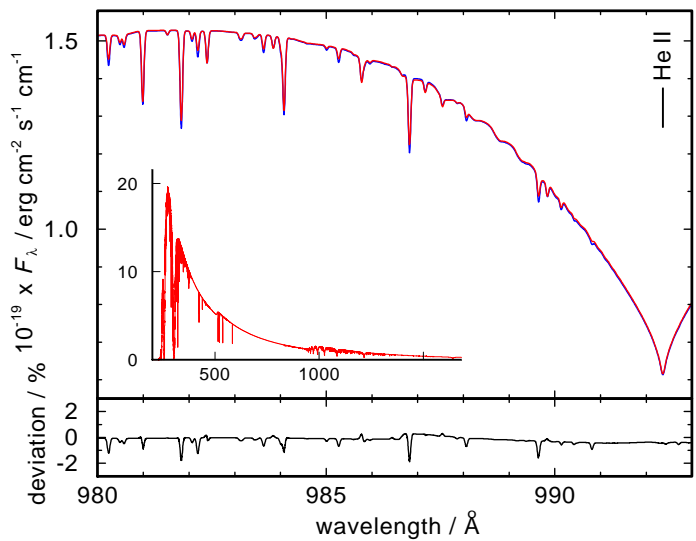}}
    \caption{Top:
             Comparison of two SEDs from He+Ga composed model-atmospheres (\Teffw{70\,000}, \loggw{7.5}, 100 times solar Ga abundance),
             on the one hand calculated with a classical (thick, blue in the online version) and 
             on the other hand with a statistical (thin, red) Ga model atom.
             The SEDs are convolved with a Gaussian (FWHM $= 0.06\,\mathrm{\AA}$) to simulate the FUSE resolution.
             The inset shows a wider wavelength range (statistical Ga model atom only).
             Bottom:
             $\left(F_\lambda^\mathrm{statistical}/F_\lambda^\mathrm{classical}\right) - 1$ illustrates the deviation between 
             the two fluxes.
            }
   \label{fig:fluxratio}
\end{figure}

Figure\,\ref{fig:fluxratio} shows a good agreement between two SEDs from models computed 
with a classical and a statistical Ga model atom. The continuum flux 
level is matched exactly, while deviations (of about 1\,\%) are visible at the line locations. 
The reason is simply the difference in the frequency grids \citep{rauchdeetjen2003}. For classical 
model atoms, all spectral lines have a very narrow frequency discretization around their centers, 
while IrOnIc uses a line-independent frequency grid for its opacity-sampling method \citep{rauchdeetjen2003}
that is equidistant in $\lambda$, for example.

For \gb and \re, TMAP uses a  frequency grid within $50\,\mathrm{\AA} \le \lambda \le 300\,000\,\mathrm{\AA}$ with
about 60\,000 frequency points.
Figure\,\ref{fig:fluxratio} (the inset) shows that the flux maximum in the model for \re is located around 
$\lambda = 300\,\mathrm{\AA}$.
In this region, we use a grid spacing of $\Delta\lambda = 5\times10^{-3}\,\mathrm{\AA}$, while around $\lambda = 1000\,\mathrm{\AA}$
(about a factor of ten below the flux maximum),
this is reduced to $\Delta\lambda = 1\times10^{-2}\,\mathrm{\AA}$ and, hence, the centers of the very narrow Ga lines do not
match perfectly. 
To calculate synthetic spectra within a restricted wavelength range,
for instance for FUSE ( $910\,\mathrm{\AA} < \lambda <  1188\,\mathrm{\AA}$)
and STIS ($1150\,\mathrm{\AA} < \lambda <  1780\,\mathrm{\AA}$), we use much finer frequency grids
that, for example, consider the centers of Ga lines in detail
(with 149\,064 and 191\,267 frequency points, respectively), and the SED agreement is even better.

For both stars, the final models of \citet{rauchetal2014ba} are adopted as start models for our calculations.
For \gb and \re, the statistics of our model atoms are summarized in Tables \ref{tab:statgb} and \ref{tab:statre},
the model abundances in Tables \ref{tab:abgb} and \ref{tab:abre}. 

\onltab{
\begin{table*}\centering
\caption{Statistics {\color{red}of} the model atoms used in our calculations for \gb.
         IG is a generic model atom that comprises Ca, Sc, V, Ti, Cr, Mn, and Co. 
         The NLTE levels for elements with an atomic number $>20$ are super levels
         build from the individual atomic levels{\color{red},} and the super lines include the 
         sample lines \citep[Kurucz LIN lines, cf\@.][]{rauchdeetjen2003}.}         
\label{tab:statgb}
\begin{tabular}{rlrrrrclrrrr}
\hline
\hline
\noalign{\smallskip}
            && \multicolumn{2}{c}{levels} & &             &&& \multicolumn{2}{c}{levels} & \\
\cline{3-4}
\cline{9-10}
\noalign{\smallskip}
\multicolumn{2}{c}{ion} & NLTE & LTE & lines  & & \multicolumn{2}{c}{ion} & NLTE & LTE & super lines & sample lines \\
\hline         
\noalign{\smallskip}

H  & \sc{i}    &  14 &   2 &    91 & & Fe & \sc{iii}  &   7 &   0 &    25 &     537\,689 \\
   & \sc{ii}   &   1 & $-$ &   $-$ & &    & \sc{iv}   &   7 &   0 &    25 &  3\,102\,371 \\
He & \sc{i}    &  45 &   0 &   121 & &    & \sc{v}    &   7 &   0 &    25 &  3\,266\,247 \\
   & \sc{ii}   &  16 &  16 &   120 & &    & \sc{vi}   &   8 &   0 &    33 &     991\,935 \\
   & \sc{iii}  &   1 & $-$ &   $-$ & &    & \sc{vii}  &   9 &   0 &    39 &     200\,455 \\
C  & \sc{ii}   &   1 &  45 &     0 & &    & \sc{viii} &   1 &   0 &     0 &              \\
   & \sc{iii}  &  58 &   9 &   329 & & Ni & \sc{iii}  &   7 &   0 &    22 &  1\,033\,920 \\
   & \sc{iv}   &  54 &   4 &   295 & &    & \sc{iv}   &   7 &   0 &    25 &  2\,512\,561 \\
   & \sc{v}    &   1 &   0 &     0 & &    & \sc{v }   &   7 &   0 &    27 &  2\,766\,664 \\
N  & \sc{ii}   &  15 & 232 &    18 & &    & \sc{vi}   &   7 &   0 &    27 &  7\,407\,763 \\
   & \sc{iii}  &  34 &  32 &   129 & &    & \sc{vii}  &   8 &   0 &    33 &  4\,195\,381 \\
   & \sc{iv}   &  90 &   4 &   546 & &    & \sc{viii} &   1 &   0 &     0 &              \\
   & \sc{v}    &  54 &   8 &   297 & & IG & \sc{iii}  &   1 &   0 &     0 &              \\
   & \sc{vi}   &   1 &   0 &     0 & &    & \sc{iv}   &   7 &   0 &    25 &  1\,581\,144 \\
O  & \sc{ii}   &   1 &  46 &     0 & &    & \sc{v}    &   7 &   0 &    23 &  2\,230\,921 \\
   & \sc{iii}  &  72 &   0 &   322 & &    & \sc{vi}   &   7 &   0 &    25 &  1\,455\,521 \\
   & \sc{iv}   &  38 &  56 &   173 & &    & \sc{vii}  &   7 &   0 &    24 &  1\,129\,512 \\
   & \sc{v}    &  76 &  50 &   472 & &    & \sc{viii} &   1 &   0 &     0 &              \\
   & \sc{vi}   &  54 &   8 &   291 & & Zn & \sc{iii}  &   1 &   0 &     0 &              \\
   & \sc{vii}  &   1 &   0 &     0 & &    & \sc{iv}   &   7 &   0 &    11 &          400 \\
Al & \sc{ii}   &   1 &   4 &     0 & &    & \sc{v}    &   7 &   0 &    15 &       1\,879 \\
   & \sc{iii}  &   7 &  29 &    10 & &    & \sc{vi}   &   1 &   0 &     0 &              \\
   & \sc{iv}   &   6 & 183 &     3 & & Ga & \sc{iii}  &   1 &   0 &     0 &              \\
   & \sc{v}    &   6 & 223 &     4 & &    & \sc{iv}   &   7 &   0 &    19 &       3\,198 \\
   & \sc{vi}   &   1 &   0 &     0 & &    & \sc{v}    &   7 &   0 &    15 &          517 \\
Si & \sc{iii}  &  17 &  17 &    28 & &    & \sc{vi}   &   7 &   0 &    13 &       1\,914 \\
   & \sc{iv}   &  16 &   7 &    44 & &    & \sc{vii}  &   1 &   0 &     0 &              \\
   & \sc{v}    &   1 &   0 &     0 & & Ge & \sc{iv}   &   1 &   0 &     0 &              \\
P  & \sc{iii}  &   3 &   7 &     0 & &    & \sc{v}    &   7 &   0 &    16 &       2\,159 \\
   & \sc{iv}   &  21 &  30 &     9 & &    & \sc{vi}   &   7 &   0 &    12 &          414 \\
   & \sc{v}    &  18 &   7 &    12 & &    & \sc{vii}  &   1 &   0 &     0 &              \\
   & \sc{vi}   &   1 &   0 &     0 & & Ba & \sc{iv}   &   1 &   0 &     0 &              \\
S  & \sc{iii}  &   1 & 230 &     0 & &    & \sc{v}    &   7 &   0 &    12 &          981 \\
   & \sc{iv}   &  17 &  83 &    32 & &    & \sc{vi}   &   7 &   0 &     6 &          162 \\
   & \sc{v}    &  39 &  71 &   107 & &    & \sc{vii}  &   7 &   0 &    11 &          493 \\
   & \sc{vi}   &  25 &  12 &    48 & &    & \sc{viii} &   1 &   0 &     0 &              \\
   & \sc{vii}  &   1 &   0 &     0 & & \multicolumn{5}{c}{}  \\
Sn & \sc{iii}  &   3 &  18 &     2 & & \multicolumn{5}{c}{}  \\
   & \sc{iv}   &   6 &   4 &     1 & & \multicolumn{5}{c}{}  \\
   & \sc{v}    &   5 &   4 &     0 & & \multicolumn{5}{c}{}  \\
   & \sc{vi}   &   6 &   0 &     0 & & \multicolumn{5}{c}{}  \\
   & \sc{vii}  &   1 &   0 &     0 & & \multicolumn{5}{c}{}  \\
\hline                                                                                     
\noalign{\smallskip}
\multicolumn{6}{r}{total}         & 17 &        78 & 1013& 1441& 4012 & 32\,424\,201 \\
\hline
\end{tabular}
\end{table*}  
}

\onltab{
\begin{table*}\centering
\caption{Same as Table\,\ref{tab:statgb}, for \re.}         
\label{tab:statre}
\begin{tabular}{rlrrrrclrrrr}
\hline
\hline
\noalign{\smallskip}
            && \multicolumn{2}{c}{levels} & &             &&& \multicolumn{2}{c}{levels} & \\
\cline{3-4}
\cline{9-10}
\noalign{\smallskip}
\multicolumn{2}{c}{ion} & NLTE & LTE & lines  & & \multicolumn{2}{c}{ion} & NLTE & LTE & super lines & sample lines \\
\hline         
\noalign{\smallskip}
He & \sc{i}    &  74 &  29 &    69 & & Fe & \sc{iii}  &   7 &   0 &    25 &      537\,689 \\
   & \sc{ii}   &  16 &  16 &   120 & &    & \sc{iv}   &   7 &   0 &    25 &   3\,102\,371 \\
   & \sc{iii}  &   1 & $-$ &   $-$ & &    & \sc{v}    &   7 &   0 &    25 &   3\,266\,247 \\
C  & \sc{iii}  &  16 & 117 &    30 & &    & \sc{vi}   &   8 &   0 &    33 &      991\,935 \\
   & \sc{iv}   &  54 &   4 &   295 & &    & \sc{vii}  &   1 &   0 &     0 &             0 \\
   & \sc{v}    &   0 &   1 &     0 & & Ni & \sc{iii}  &   7 &   0 &    22 &   1\,033\,920 \\
N  & \sc{iii}  &  18 &  48 &    42 & &    & \sc{iv}   &   7 &   0 &    25 &   2\,512\,561 \\
   & \sc{iv}   &  48 &  46 &   201 & &    & \sc{v}    &   7 &   0 &    27 &   2\,766\,664 \\
   & \sc{v}    &  35 &  27 &   149 & &    & \sc{vi}   &   7 &   0 &    27 &   7\,407\,763 \\
   & \sc{vi}   &   0 &   1 &     0 & &    & \sc{vii}  &   1 &   0 &     0 &             0 \\
O  & \sc{iii}  &   1 &   1 &     0 & & IG & \sc{iii}  &   7 &   0 &    24 &   4\,963\,441 \\
   & \sc{iv}   &  69 &  28 &   432 & &    & \sc{iv}   &   7 &   0 &    25 &   1\,581\,144 \\
   & \sc{v}    &  90 &  44 &   610 & &    & \sc{v}    &   7 &   0 &    23 &   2\,230\,921 \\
   & \sc{vi}   &  14 &  48 &    33 & &    & \sc{vi}   &   7 &   0 &    25 &   1\,455\,521 \\
   & \sc{viii} &   0 &   1 &     0 & &    & \sc{vii}  &   1 &   0 &     0 &             0 \\
Si & \sc{iii}  &   1 &  33 &     0 & & Zn & \sc{iii}  &   1 &   0 &     0 &             0 \\
   & \sc{iv}   &  16 &   7 &    44 & &    & \sc{iv}   &   7 &   0 &    11 &           400 \\
   & \sc{v}    &   1 &   0 &     0 & &    & \sc{v}    &   7 &   0 &    15 &        1\,879 \\
P  & \sc{iv}   &   1 &  50 &     0 & &    & \sc{vi}   &   1 &   0 &     0 &             0 \\
   & \sc{v}    &  18 &   7 &    12 & & Ga & \sc{iii}  &   1 &   0 &     0 &             0 \\
   & \sc{vi}   &   1 &   0 &     0 & &    & \sc{iv}   &   7 &   0 &    19 &        3\,198 \\
S  & \sc{iv}   &   1 &  99 &     0 & &    & \sc{v}    &   7 &   0 &    15 &           517 \\
   & \sc{v}    &  23 &  87 &    47 & &    & \sc{vi}   &   7 &   0 &    13 &        1\,914 \\
   & \sc{vi}   &  12 &  25 &    25 & &    & \sc{vii}  &   1 &   0 &     0 &             0 \\
   & \sc{vii}  &   1 &   1 &     0 & & Ge & \sc{iv}   &   1 &   0 &     0 &             0 \\
\multicolumn{5}{c}{}               & &    & \sc{v}    &   7 &   0 &    16 &        2\,159 \\
\multicolumn{5}{c}{}               & &    & \sc{vi}   &   7 &   0 &    12 &           414 \\
\multicolumn{5}{c}{}               & &    & \sc{vii}  &   1 &   0 &     0 &             0 \\
\multicolumn{5}{c}{}               & & Kr & \sc{iii}  &   1 &  25 &     0 &             0 \\
\multicolumn{5}{c}{}               & &    & \sc{iv}   &  38 &   0 &     0 &             0 \\
\multicolumn{5}{c}{}               & &    & \sc{v}    &  25 &   0 &     0 &             0 \\
\multicolumn{5}{c}{}               & &    & \sc{vi}   &  46 &   0 &   887 &             0 \\
\multicolumn{5}{c}{}               & &    & \sc{vii}  &  14 &   0 &    37 &             0 \\
\multicolumn{5}{c}{}               & &    & \sc{viii} &   1 &   0 &     0 &             0 \\
\multicolumn{5}{c}{}               & & Xe & \sc{iii}  &  11 &  11 &     0 &             0 \\
\multicolumn{5}{c}{}               & &    & \sc{iv}   &  44 &   0 &     0 &             0 \\
\multicolumn{5}{c}{}               & &    & \sc{v}    &  29 &   0 &     0 &             0 \\
\multicolumn{5}{c}{}               & &    & \sc{vi}   &  82 &   0 &   887 &             0 \\
\multicolumn{5}{c}{}               & &    & \sc{vii}  &  86 &   3 &     2 &             0 \\
\multicolumn{5}{c}{}               & &    & \sc{viii} &   1 &   0 &     0 &             0 \\
\multicolumn{5}{c}{}               & & Ba & \sc{iv}   &   1 &   0 &     0 &             0 \\
\multicolumn{5}{c}{}               & &    & \sc{v}    &   7 &   0 &    12 &           981 \\
\multicolumn{5}{c}{}               & &    & \sc{vi}   &   7 &   0 &     6 &           162 \\
\multicolumn{5}{c}{}               & &    & \sc{vii}  &   7 &   0 &    11 &           493 \\
\multicolumn{5}{c}{}               & &    & \sc{viii} &   1 &   0 &     0 &             0 \\
\hline                                                                                     
\noalign{\smallskip}
\multicolumn{6}{r}{total}            & 16 &        70 & 953 & 870 &  4358 &  31\,862\,294 \\
\hline
\end{tabular}
\end{table*}  
}

\begin{table}\centering
  \caption{Photospheric abundances of \gb as used in our final model.
           [X] denotes log (abundance/solar abundance) of species X.}         
\label{tab:abgb}
\setlength{\tabcolsep}{.4em}
\begin{tabular}{rlr@{.}lr@{.}lr@{.}l}
\hline
\hline
\noalign{\smallskip}                                                                                          
&                          & \multicolumn{2}{c}{mass}   & \multicolumn{2}{c}{number}  & \multicolumn{2}{c}{}                      \\
\cline{3-6}                     
\multicolumn{8}{c}{}                                                                                                 \vspace{-5mm}\\
& element                  & \multicolumn{2}{c}{}       & \multicolumn{2}{c}{}        & \multicolumn{2}{c}{~~~~~[X]} \vspace{-2mm}\\
&                          & \multicolumn{4}{c}{fraction}                             & \multicolumn{2}{c}{}                      \\
\cline{2-8}                     
\noalign{\smallskip}                                                                                   
\smspr & \mmspr H                        & $ 9$&$99\times 10^{-1}$ & $ 0$&$99998$          & $  0$&$132$ \\
       & \mmspr He\tablefootmark{a}      & $ 1$&$98\times 10^{-5}$ & $ 5$&$0\times 10^{-6}$ & $ -4$&$099$ \\
       & \mmspr C                        & $ 6$&$31\times 10^{-6}$ & $ 5$&$3\times 10^{-7}$ & $ -2$&$574$ \\
       & \mmspr N                        & $ 2$&$08\times 10^{-6}$ & $ 1$&$5\times 10^{-7}$ & $ -2$&$522$ \\
       & \mmspr O                        & $ 1$&$90\times 10^{-5}$ & $ 1$&$2\times 10^{-6}$ & $ -2$&$479$ \\
       & \mmspr Al                       & $ 1$&$12\times 10^{-5}$ & $ 4$&$2\times 10^{-7}$ & $ -0$&$695$ \\
       & \mmspr Si                       & $ 5$&$29\times 10^{-5}$ & $ 1$&$9\times 10^{-6}$ & $ -1$&$099$ \\
       & \mmspr P                        & $ 1$&$54\times 10^{-6}$ & $ 5$&$0\times 10^{-8}$ & $ -0$&$579$ \\
       & \mmspr S                        & $ 5$&$72\times 10^{-6}$ & $ 1$&$8\times 10^{-7}$ & $ -1$&$733$ \\
       & \mmspr IG\tablefootmark{b}      & $ 1$&$78\times 10^{-6}$ & $ 4$&$0\times 10^{-8}$ & $ -1$&$558$ \\
       & \mmspr Fe                       & $ 6$&$50\times 10^{-4}$ & $ 1$&$2\times 10^{-5}$ & $ -0$&$298$ \\
       & \mmspr Ni                       & $ 3$&$84\times 10^{-5}$ & $ 6$&$6\times 10^{-7}$ & $ -0$&$269$ \\
       & \mmspr Zn                       & $ 3$&$50\times 10^{-6}$ & $ 5$&$4\times 10^{-8}$ & $  0$&$304$ \\
       & \mmspr Ga                       & $ 2$&$00\times 10^{-6}$ & $ 3$&$0\times 10^{-8}$ & $  1$&$560$ \\
       & \mmspr Ge                       & $ 3$&$24\times 10^{-6}$ & $ 4$&$5\times 10^{-8}$ & $  1$&$135$ \\
       & \mmspr Sn                       & $ 3$&$53\times 10^{-7}$ & $ 3$&$0\times 10^{-9}$ & $  2$&$588$ \\
       & \mmspr Ba                       & $ 4$&$00\times 10^{-6}$ & $ 2$&$9\times 10^{-8}$ & $  3$&$495$ \\
\hline
\end{tabular}
\tablefoot{~\\
\tablefoottext{a}{Upper limit given by \citet{rauchetal2013}.}\\
\tablefoottext{b}{Generic model atom \citep{rauchdeetjen2003} that comprises Ca, Sc, V, Ti, Cr, Mn, and Co, constructed with
                  a relative abundance pattern that uses the upper abundance limits of Ti, Cr, Mn, and Co given by \citet{rauchetal2013}
                  and $1.0\,\times\,10^{-7}$ (mass fraction) for Ca, Sc, and V.}           
}
\end{table}

\begin{table}\centering 
  \caption{Same as Table\,\ref{tab:abgb}, for \re.}
\label{tab:abre}
\setlength{\tabcolsep}{.4em}
\begin{tabular}{rlr@{.}lr@{.}lr@{.}l}
\hline
\hline
\noalign{\smallskip}                                                                                          
&                          & \multicolumn{2}{c}{mass}   & \multicolumn{2}{c}{number}  & \multicolumn{2}{c}{}                      \\
\cline{3-6}                     
\multicolumn{8}{c}{}                                                                                                 \vspace{-5mm}\\
& element                  & \multicolumn{2}{c}{}       & \multicolumn{2}{c}{}        & \multicolumn{2}{c}{~~~~~[X]} \vspace{-2mm}\\
&                          & \multicolumn{4}{c}{fraction}                             & \multicolumn{2}{c}{}                      \\
\cline{2-8}                     
\noalign{\smallskip}                                                                                   
\smspr & \mmspr He                      & $ 9$&$75\times 10^{-1}$ & $ 9$&$9\times 10^{-1}$ & $  0$&$592$ \\
       & \mmspr C                       & $ 2$&$23\times 10^{-2}$ & $ 7$&$6\times 10^{-3}$ & $  0$&$974$ \\
       & \mmspr N                       & $ 1$&$73\times 10^{-4}$ & $ 5$&$0\times 10^{-5}$ & $ -0$&$602$ \\
       & \mmspr O                       & $ 1$&$97\times 10^{-3}$ & $ 5$&$0\times 10^{-4}$ & $ -0$&$464$ \\
       & \mmspr Si                      & $ 1$&$61\times 10^{-4}$ & $ 2$&$3\times 10^{-5}$ & $ -0$&$617$ \\
       & \mmspr P                       & $ 1$&$15\times 10^{-6}$ & $ 1$&$5\times 10^{-7}$ & $ -0$&$705$ \\
       & \mmspr S                       & $ 3$&$97\times 10^{-5}$ & $ 5$&$0\times 10^{-6}$ & $ -0$&$892$ \\
       & \mmspr IG                      & $ 1$&$00\times 10^{-6}$ & $ 9$&$1\times 10^{-8}$ & $ -1$&$807$ \\
       & \mmspr Fe                      & $ 1$&$30\times 10^{-5}$ & $ 9$&$5\times 10^{-7}$ & $ -1$&$997$ \\
       & \mmspr Ni                      & $ 7$&$26\times 10^{-5}$ & $ 5$&$0\times 10^{-6}$ & $  0$&$008$ \\
       & \mmspr Zn                      & $ 1$&$13\times 10^{-4}$ & $ 7$&$1\times 10^{-6}$ & $  1$&$814$ \\
       & \mmspr Ga                      & $ 3$&$45\times 10^{-5}$ & $ 2$&$0\times 10^{-6}$ & $  2$&$790$ \\
       & \mmspr Ge                      & $ 1$&$59\times 10^{-4}$ & $ 8$&$9\times 10^{-6}$ & $  2$&$825$ \\
       & \mmspr Kr                      & $ 5$&$05\times 10^{-5}$ & $ 2$&$5\times 10^{-6}$ & $  2$&$666$ \\
       & \mmspr Xe                      & $ 6$&$30\times 10^{-5}$ & $ 1$&$9\times 10^{-6}$ & $  4$&$732$ \\
       & \mmspr Ba                      & $ 3$&$58\times 10^{-4}$ & $ 1$&$1\times 10^{-5}$ & $  5$&$447$ \\
\hline
\end{tabular}
\end{table}

The SEDs calculated for this analysis are available via
the registered Theoretical Stellar Spectra Access 
(TheoSSA\footnote{\url{http://dc.g-vo.org/theossa}}) Virtual Observatory (VO) service.

\section{Photospheric Ga abundances in \gb and \re}
\label{sect:abund}

\begin{figure*}
   \resizebox{\hsize}{!}{\includegraphics{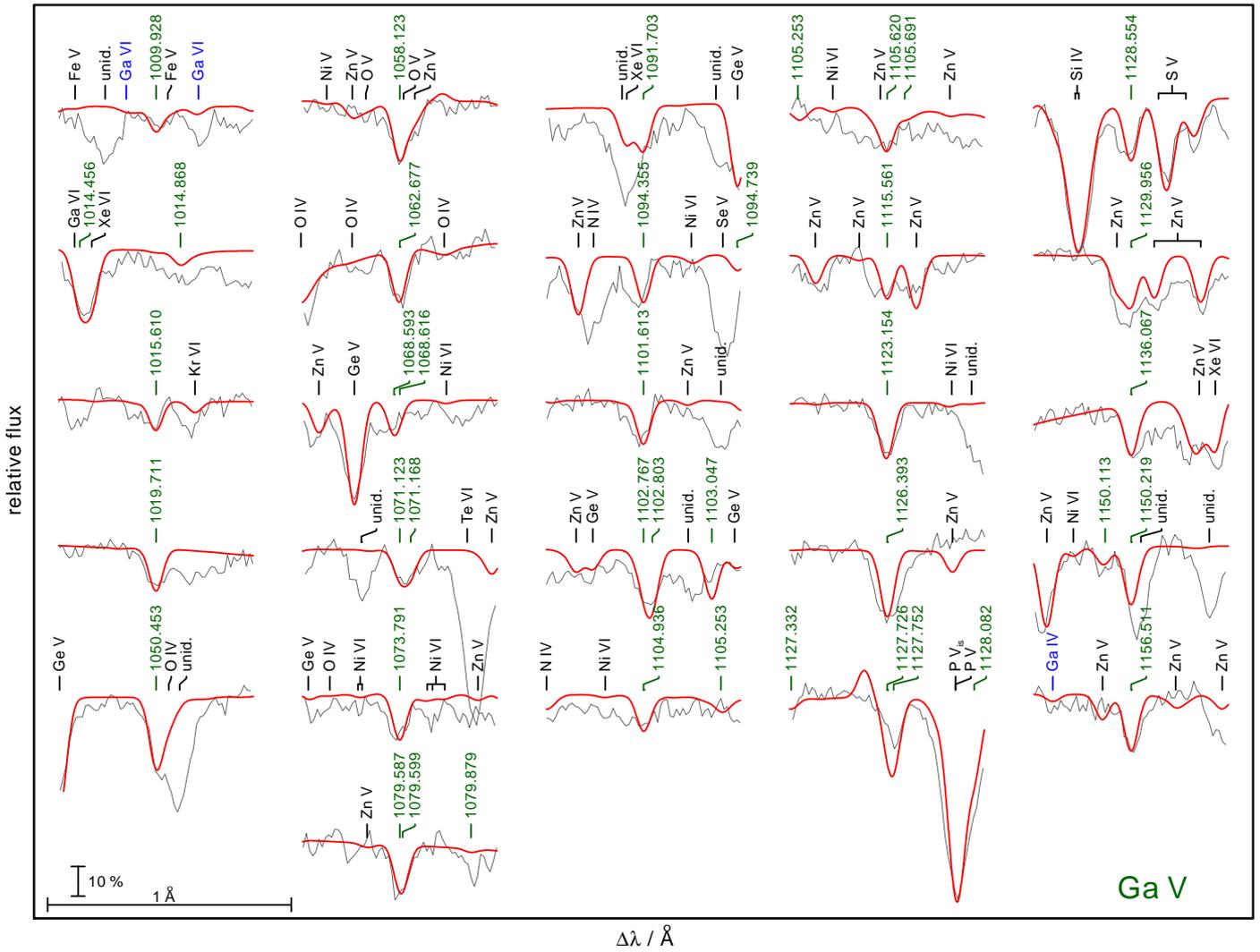}}
    \caption{Strongest \ion{Ga}{v} lines in the FUSE observation of \re, 
             labeled (green in the online version) with their
             wavelengths from Table\,\ref{tab:gav:loggf}.
             The vertical bar indicates 10\,\% of the continuum flux.
             Identified lines are marked. ``unid.'' denotes unidentified, ``is'' interstellar.
            }
   \label{fig:gafuse}
\end{figure*}

At a Ga abundance of $3.5\,\times\,10^{-5}$ (mass fraction, 625 times the solar value), 
we were able to reproduce the identified \ion{Ga}{v}
lines (Table\,\ref{tab:lineids}) in the FUSE spectrum of \re (Fig.\,\ref{fig:gafuse}).

\begin{figure}
   \resizebox{\hsize}{!}{\includegraphics{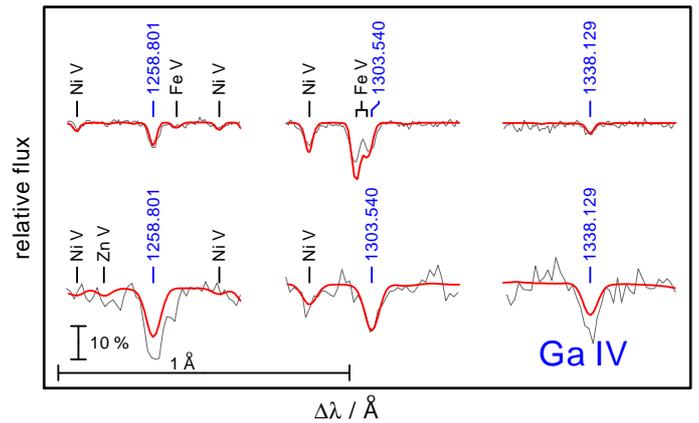}}
    \caption{Strongest \ion{Ga}{iv} lines
             in the STIS observations of \gb (top) and \re (bottom),
             labeled (blue in the online version) with their
             wavelengths from Table\,\ref{tab:gaiv:loggf}.
             Identified lines of other species are marked.
            }
   \label{fig:gastis}
\end{figure}

\begin{figure}
   \resizebox{\hsize}{!}{\includegraphics{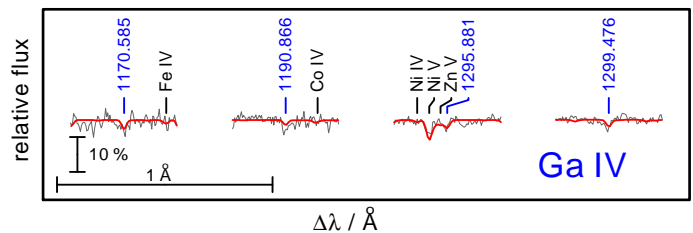}}
    \caption{Weaker \ion{Ga}{iv} lines
             in the STIS observation of \gb,
             labeled (blue in the online version) with their
             wavelengths from Table\,\ref{tab:gaiv:loggf}.
             Identified lines of other species are marked.
            }
   \label{fig:gaaddstis}
\end{figure}

The search for Ga lines in the FUSE observation of \gb was entirely negative.
Our models for \gb showed that the strongest \ion{Ga}{iv} lines given by \citet[with intensities of 1000]{shiraietal2007}
are located in the STIS wavelength range. These are
\ionw{Ga}{iv}{1258.801}  (3d$^9$4s $^3$D$_3$ - 3d$^9$4p $^3$F$^\mathrm{o}_4$),
\ionw{Ga}{iv}{1303.540}  (3d$^9$4s $^3$D$_2$ - 3d$^9$4p $^3$F$^\mathrm{o}_3$), and
\ionw{Ga}{iv}{1338.129}  (3d$^9$4s $^3$D$_3$ - 3d$^9$4p $^3$P$^\mathrm{o}_2$). They have the largest
equivalent widths of all Ga lines in our models for \gb and are positively identified in the 
STIS observation (Fig.\,\ref{fig:gastis}). \ionw{Ga}{iv}{1303.540} is a blend with \ion{Fe}{v} lines and, 
thus, uncertain, while the other two lines are isolated.
We can reproduce their observed strengths at an abundance of 
$2.0\,\times\,10^{-6}$ (by mass, 37 times solar, Fig.\,\ref{fig:gastis}).
At this abundance, four of the next strongest \ion{Ga}{iv} lines 
\citep[\ionww{Ga}{iv}{1170.585, 1190.866, 1295.881, 1299.476}, with intensities of 220, 220, 650, and 600, respectively,][]{shiraietal2007}
can also be identified and modeled, but they appear just above the noise level (Fig.\,\ref{fig:gaaddstis}). 
However, no detectable Ga line is predicted by our models in the FUSE wavelength range.

\ionww{Ga}{iv}{1258.801, 1303.540, 1338.129} are also visible in the STIS observation of \re 
(Fig.\,\ref{fig:gastis}) and agree well with the Ga abundance ($3.5\,\times\,10^{-5}$ by mass) 
determined from the FUSE spectrum.

\section{Results and conclusions}
\label{sect:results}

We have unambiguously identified \ion{Ga}{v} lines in the observed high-resolution UV spectra of \gb 
and \re. These lines are well reproduced by our NLTE model-atmosphere calculations using
our newly calculated \ion{Ga}{iv-vi} oscillator strengths.

\begin{figure}
   \resizebox{\hsize}{!}{\includegraphics{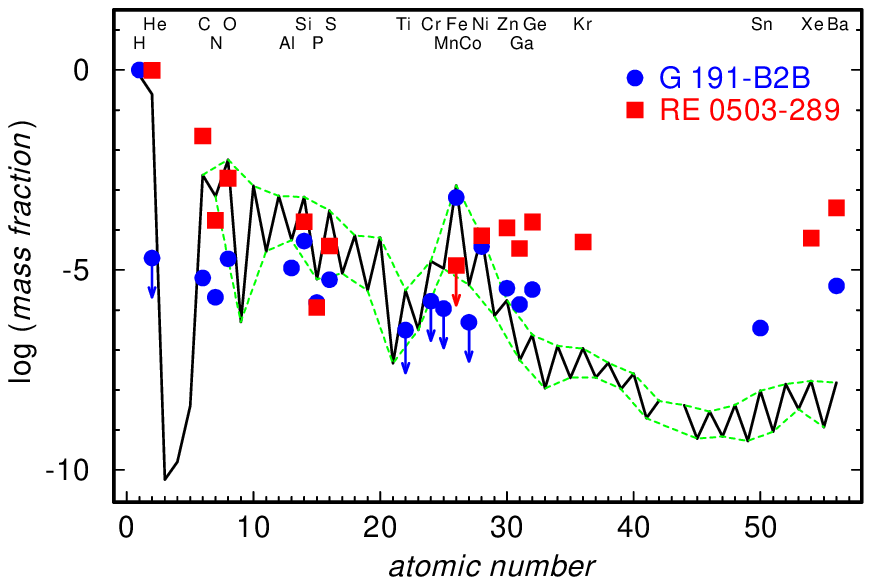}}
    \caption{Solar abundances \citep[thick line; the dashed lines
             connect the elements with even and with odd atomic number]{asplundetal2009}
             compared with the determined photospheric abundances of 
             \gb \citep[blue circles,][]{rauchetal2013} and 
             \re \citep[red squares,][and this work]{dreizlerwerner1996,werneretal2012,rauchetal2013,rauchetal2014zn,rauchetal2014ba}.
             The uncertainties of the WD abundances are about 0.2\,dex in general. Arrows indicate upper limits.
            }
   \label{fig:X}
\end{figure}

We determined photospheric abundances of 
$\log\,\mathrm{Ga} = -5.69 \pm 0.2$ (mass fraction, $1.7 - 2.5\,\times\,10^{-6}$,  30 --  44 times the solar abundance) and
$\log\,\mathrm{Ga} = -4.49 \pm 0.1$                ($3.0 - 4.0\,\times\,10^{-5}$, 536 -- 715 times     solar)
for 
the DA-type white dwarf \gb and 
the DO-type white dwarf \re, respectively.
These highly supersolar Ga abundances agree with the high abundances of other trans-iron elements in \gb and \re
(Fig.\,\ref{fig:X}).

The \ion{Ga}{iv}\,/\,\ion{Ga}{v} ionization balance is well reproduced in \re (Figs.\,\ref{fig:gastis}, \ref{fig:gaaddstis}).

The identification of lines of Ga and its precise abundance determination became possible 
only because reliable transition probabilities for \ion{Ga}{iv}, \ion{Ga}{v}, and \ion{Ga}{vi}
were computed.
Analogous calculations for other highly ionized trans-iron elements are highly desirable.
The precise measurement of their spectra, that is, their line wavelengths and relative strengths,
as well as the determination of level energies and the calculation of transition probabilities 
remains a challenge for atomic and theoretical physicists.

\begin{acknowledgements}
TR is supported by the German Aerospace Center (DLR, grant 05\,OR\,1402).
Financial support from the Belgian FRS-FNRS is also acknowledged. 
PQ is research director of this organization.
This research has made use of the SIMBAD database, operated at CDS, Strasbourg, France.
Some of the data presented in this paper were obtained from the
Mikulski Archive for Space Telescopes (MAST). STScI is operated by the
Association of Universities for Research in Astronomy, Inc., under NASA
contract NAS5-26555. Support for MAST for non-HST data is provided by
the NASA Office of Space Science via grant NNX09AF08G and by other
grants and contracts. 

\end{acknowledgements}

\bibliographystyle{aa}
\bibliography{25326}

\onecolumn
\begin{longtable}{rlrcl}
\caption{\label{tab:lineids}Strongest Ga lines in the stellar-atmosphere models of \re.
                            The given wavelengths correspond to those given in 
                            Tables \ref{tab:gaiv:loggf}, \ref{tab:gav:loggf}, and \ref{tab:gavi:loggf}.
                            Column 3 gives the equivalent widths of the computed lines.
                            Column 4 denotes ``$-$'' = not observed, ``+'' = observed, ``:'' = uncertain.} \\
\hline\hline
\noalign{\smallskip}
wavelength / \AA & ion & $W_\mathrm{\lambda}\,/\,\mathrm{m\AA}$ & id & comment\\
\noalign{\smallskip}
\hline
\endfirsthead
\caption{continued.}\\
\hline\hline
\noalign{\smallskip}
wavelength / \AA & ion & $W_\mathrm{\lambda}\,/\,\mathrm{m\AA}$ & id & comment \\
\noalign{\smallskip}
\hline
\noalign{\smallskip}
\endhead
\hline
\noalign{\smallskip}
\endfoot
\noalign{\smallskip}
 953.738                  &  \ion{Ga}{vi}              & $   0.97 $ &$-$&                                                   \\
 965.237  965.272         &  \ion{Ga}{iv}              & $   2.23 $ & : & blend                                             \\
 979.383                  &  \ion{Ga}{vi}              & $   0.99 $ &$-$&                                                   \\
 979.614                  &  \ion{Ga}{v}               & $   3.65 $ & : &                                                   \\
 980.240                  &  \ion{Ga}{vi}              & $   1.07 $ & : &                                                   \\
 980.988                  &  \ion{Ga}{v}               & $   2.88 $ &$-$&                                                   \\
 981.831                  &  \ion{Ga}{iv}              & $   1.57 $ & : &                                                   \\
 984.078                  &  \ion{Ga}{v}               & $   3.09 $ & : &                                                   \\
1002.985                  &  \ion{Ga}{vi}              & $   1.49 $ & : &                                                   \\
1004.170                  &  \ion{Ga}{vi}              & $   1.15 $ &$-$& blend \ionw{Ba}{v}{1004.093}                      \\
1006.396                  &  \ion{Ga}{vi}              & $   2.55 $ &$-$&                                                   \\
1006.894                  &  \ion{Ga}{vi}              & $   1.54 $ & : &                                                   \\
1008.924                  &  \ion{Ga}{vi}              & $   2.44 $ & : &                                                   \\
1009.512                  &  \ion{Ga}{vi}              & $   1.03 $ & : &                                                   \\
1009.928                  &  \ion{Ga}{v}               & $   6.07 $ & + &                                                   \\
1010.102                  &  \ion{Ga}{vi}              & $   1.88 $ & : &                                                   \\
1011.047                  &  \ion{Ga}{vi}              & $   2.29 $ & : &                                                   \\
1014.434 1014.456         &  \ion{Ga}{vi} \ion{Ga}{v}  & $  12.34 $ & : & blend \ionw{Xe}{vi}{1014.505}                     \\
1014.868                  &  \ion{Ga}{v}               & $   2.19 $ & : & blend \ionw{O}{iv}{1014.90}                       \\
1015.598 1015.610         &  \ion{Ga}{vi} \ion{Ga}{v}  & $   6.19 $ & + & blend                                             \\
1019.711                  &  \ion{Ga}{v}               & $   8.30 $ & + &                                                   \\
1032.375                  &  \ion{Ga}{v}               & $   2.51 $ &$-$&                                                   \\
1033.549                  &  \ion{Ga}{v}               & $   3.83 $ & : &                                                   \\
1034.822                  &  \ion{Ga}{v}               & $   1.15 $ &$-$& blend \ionw{Xe}{vi}{1034.784}                     \\
1038.778                  &  \ion{Ga}{v}               & $   5.12 $ & : &                                                   \\
1040.204                  &  \ion{Ga}{v}               & $   1.79 $ & : &                                                   \\
1045.850                  &  \ion{Ga}{v}               & $   3.36 $ & : &                                                   \\
1047.504                  &  \ion{Ga}{v}               & $   5.08 $ & : & blend \ionw{O}{iv}{1047.590}                      \\
1050.453                  &  \ion{Ga}{v}               & $  16.83 $ & + & blend \ionw{O}{iv}{1050.505}                      \\
1054.430                  &  \ion{Ga}{v}               & $   1.70 $ &$-$&                                                   \\
1054.563                  &  \ion{Ga}{v}               & $   5.07 $ & : & blend \ionw{Ge}{v}{1054.590}                      \\
1058.123                  &  \ion{Ga}{v}               & $  14.02 $ & + & blend \ionw{Zn}{v}{1058.190}                      \\
1062.677                  &  \ion{Ga}{v}               & $  10.84 $ & + & blend \ionw{Ge}{v}{1062.574}                      \\
1063.807                  &  \ion{Ga}{v}               & $   3.61 $ & : &                                                   \\
1065.371                  &  \ion{Ga}{v}               & $   1.34 $ &$-$&                                                   \\
1066.724                  &  \ion{Ga}{v}               & $   8.25 $ & : & blend \ionww{Si}{iv}{1066.636, 1066.650}          \\
1068.593 1068.616         &  \ion{Ga}{v}               & $   7.90 $ & + & blend                                             \\
1069.484 1069.530 1069.587&  \ion{Ga}{v}               & $  15.77 $ & : & blend                                             \\
1071.123 1071.168         &  \ion{Ga}{v}               & $  11.28 $ & + & blend                                             \\
1073.791 1073.814         &  \ion{Ga}{v}  \ion{Ga}{vi} & $  11.12 $ & + & blend                                             \\
1074.911 1074.966         &  \ion{Ga}{v}  \ion{Ga}{iv} & $   1.12 $ & : & blend                                             \\
1078.225                  &  \ion{Ga}{v}               & $   2.31 $ & : &                                                   \\
1079.587 1079.599         &  \ion{Ga}{v}               & $  11.47 $ & + & blend                                             \\
1080.474                  &  \ion{Ga}{v}               & $   1.84 $ & : &                                                   \\
1080.988                  &  \ion{Ga}{v}               & $   3.72 $ & : & blend \ionw{Xe}{vi}{1080.869}                     \\
1087.358                  &  \ion{Ga}{v}               & $   2.64 $ &$-$& blend (unknown)                                   \\
1088.068                  &  \ion{Ga}{v}               & $   1.27 $ &$-$&                                                   \\
1091.703                  &  \ion{Ga}{v}               & $   9.56 $ & + & blend \ionw{Xe}{vi}{1091.634}                     \\
1094.355                  &  \ion{Ga}{v}               & $  11.50 $ & + &                                                   \\
1094.739                  &  \ion{Ga}{v}               & $   3.16 $ &$-$& blend \ionw{Se}{v}{1094.68}                       \\
1095.110                  &  \ion{Ga}{v}               & $  11.66 $ & : & blend                                             \\
1100.401                  &  \ion{Ga}{v}               & $   7.83 $ & : &                                                   \\
1101.613                  &  \ion{Ga}{v}               & $   9.68 $ & + &                                                   \\
1102.767 1102.803         &  \ion{Ga}{v}               & $  20.30 $ & + & blend                                             \\
1103.047                  &  \ion{Ga}{v}               & $  12.37 $ & : &                                                   \\
1104.936                  &  \ion{Ga}{v}               & $   8.42 $ & + &                                                   \\
1105.253                  &  \ion{Ga}{v}               & $   2.33 $ &$-$&                                                   \\
1105.620                  &  \ion{Ga}{v}               & $   8.47 $ & + &                                                   \\
1107.763                  &  \ion{Ga}{v}               & $   0.86 $ &$-$& blend \ionww{C}{iv}{1107.591. 1107.930, 1107.979} \\
1109.829                  &  \ion{Ga}{v}               & $   1.45 $ &$-$&                                                   \\
1115.561                  &  \ion{Ga}{v}               & $   9.91 $ & + &                                                   \\
1118.018                  &  \ion{Ga}{v}               & $   1.23 $ &$-$& blend \ionw{P}{v}{1117.976}                       \\
1118.318                  &  \ion{Ga}{v}               & $  10.79 $ & : &                                                   \\
1120.260                  &  \ion{Ga}{v}               & $   2.80 $ &$-$& blend \ionw{Zn}{v}{1120.325}                      \\
1123.154                  &  \ion{Ga}{v}               & $  11.90 $ & + &                                                   \\
1123.646                  &  \ion{Ga}{v}               & $  10.79 $ & : & blend (unknown)                                   \\
1126.393                  &  \ion{Ga}{v}               & $  17.59 $ & + & blend (two \ion{Ga}{v} lines, same $\lambda$)     \\
1127.332                  &  \ion{Ga}{v}               & $   2.42 $ & : &                                                   \\
1127.726 1127.752         &  \ion{Ga}{v}               & $  20.80 $ & + & blend                                             \\
1128.082                  &  \ion{Ga}{v}               & $  13.16 $ & : & blend \ionw{P}{v}{1128.000}                       \\
1128.554                  &  \ion{Ga}{v}               & $  14.86 $ & + & blend \ionww{S}{v}{1128.666, 1128.779}            \\
1129.956                  &  \ion{Ga}{v}               & $  12.35 $ & + &                                                   \\              
1131.452                  &  \ion{Ga}{v}               & $  10.79 $ & : & blend (unknown)                                   \\
1132.054                  &  \ion{Ga}{v}               & $   4.38 $ &$-$&                                                   \\
1132.157                  &  \ion{Ga}{v}               & $   3.04 $ &$-$&                                                   \\
1133.247                  &  \ion{Ga}{v}               & $   4.08 $ &$-$&                                                   \\
1133.903                  &  \ion{Ga}{v}               & $   3.99 $ & : &                                                   \\
1136.067                  &  \ion{Ga}{v}               & $  12.35 $ & + &                                                   \\
1138.187                  &  \ion{Ga}{v}               & $   7.88 $ & : & blend \ionw{Zn}{v}{1138.248}                      \\
1143.367                  &  \ion{Ga}{v}               & $   3.02 $ & : & blend \ionww{Zn}{v}{1143.304, 1143.403}           \\
1145.974                  &  \ion{Ga}{v}               & $   2.05 $ &$-$&                                                   \\
1148.409                  &  \ion{Ga}{v}               & $   3.65 $ & : &                                                   \\
1150.113                  &  \ion{Ga}{v}               & $   3.31 $ & + &                                                   \\
1150.219                  &  \ion{Ga}{v}               & $  16.35 $ & + &                                                   \\
1155.976                  &  \ion{Ga}{v}               & $   2.19 $ &$-$&                                                   \\
1156.190                  &  \ion{Ga}{iv}              & $   1.94 $ &$-$& blend \ionw{Zn}{v}{1156.394}                      \\
1156.511                  &  \ion{Ga}{v}               & $  13.04 $ & + & blend \ionw{Zn}{v}{1156.520}                      \\
1157.729                  &  \ion{Ga}{v}               & $   2.43 $ & : &                                                   \\
1158.534                  &  \ion{Ga}{v}               & $   3.83 $ &$-$& blend \ionw{Zn}{v}{1158.476}                      \\
1160.847                  &  \ion{Ga}{v}               & $   1.28 $ & : &                                                   \\
1163.609                  &  \ion{Ga}{iv}              & $   1.77 $ &$-$&                                                   \\
1170.585                  &  \ion{Ga}{iv}              & $   3.32 $ & : &                                                   \\
1178.795                  &  \ion{Ga}{v}               & $   4.22 $ & : & blend \ionw{Zn}{v}{1078.808}                      \\
1183.110                  &  \ion{Ga}{v}               & $   1.45 $ &$-$& blend \ionw{Zn}{v}{1183.024}                      \\
1183.656                  &  \ion{Ga}{v}               & $   3.00 $ &$-$&                                                   \\
1185.226                  &  \ion{Ga}{iv}              & $   2.00 $ &$-$&                                                   \\
\hline
\end{longtable}
\twocolumn

\end{document}